\documentclass[runningheads]{llncs}

\usepackage{amsmath,amssymb}
\usepackage{graphicx}
\usepackage{subcaption}
\usepackage{cite}
\usepackage{pifont}
\usepackage{booktabs}
\usepackage{url}
\newcommand{\cmark}{\ding{51}}
\newcommand{\xmark}{\ding{55}}

\begin{document}

\title{Composite Reward Design in PPO-Driven Adaptive Filtering}
\titlerunning{PPO-Driven Adaptive Filtering}

\author{A.~Burkan~Bereketoglu\inst{1}\orcidID{0000-0001-5034-4125}}
\authorrunning{A.~B.~Bereketoglu}

\institute{
Department of Informatics, University of Sussex, Brighton BN1 9QJ, UK\\
\email{a.bereketoglu@sussex.ac.uk}
}

\maketitle

\begin{abstract}
Model-free and reinforcement learning-based adaptive filtering methods are gaining traction for denoising in dynamic, non-stationary environments such as wireless signal channels, biomedical monitoring, and sensor networks. Traditional filters such as LMS, RLS, Wiener, and Kalman are often limited by assumptions of stationarity, the need for exact noise statistics, or fragile parameter tuning. This paper proposes an adaptive filtering framework using Proximal Policy Optimization (PPO), guided by a composite reward that balances SNR improvement, MSE reduction, and residual smoothness. We frame adaptive filtering as a Markov decision process and train a PPO agent to adjust filter coefficients directly in response to changing noise. Experiments on synthetic nonstationary signals with diverse noise types show that the PPO agent generalizes beyond its training distribution. Moreover, real-world analysis is made and evaluated on ECG recordings from the MIT-BIH Noise Stress Test Database corrupted by baseline wander, electrode motion, and muscle artifacts. The learned PPO policy achieves real-time inference and slightly outperforms strong classical baselines on ECG denoising. These results demonstrate the viability of policy-gradient reinforcement learning as a computationally efficient and flexible tool for adaptive filtering in nonlinear, time-varying dynamical systems.
\keywords{Reinforcement learning \and Adaptive filtering \and Noise reduction \and PPO}
\end{abstract}

\section{Introduction}
Wireless communication systems, biomedical devices, and sensor networks often operate in noisy, time-varying environments where effective denoising is critical. Adaptive filters such as LMS~\cite{widrow85}, RLS~\cite{haykin13}, Wiener~\cite{haykin13}, and Kalman filters~\cite{kalman60} are widely used, but they can struggle in highly dynamic scenarios due to rigid assumptions. LMS requires careful step-size tuning and assumes relatively stationary noise; RLS is memory-intensive and may become unstable under impulsive interference; Wiener filtering is optimal only under stationary assumptions with known statistics; and Kalman filters demand accurate state-space models and noise covariances. In practice, classical filters often cannot fully cope with rapidly shifting or uncertain noise conditions without manual re-calibration.

Reinforcement learning (RL) offers a data-driven alternative for adaptive signal filtering. Instead of relying on a fixed model, an RL agent learns to adjust filter parameters through interaction with the environment and feedback through a reward signal. Unlike
supervised approaches, RL does not require ground-truth clean signals at run-time and can continually adapt online to new noise conditions.  Policy-gradient methods such as Proximal Policy Optimization (PPO)~\cite{schulman17} are particularly suitable because they support continuous actions and use conservative clipped updates that improve training stability. PPO uses a clipped surrogate objective to stabilize training, addressing the risk of divergence in high-variance environments, and naturally handles continuous action spaces (e.g., filter weight adjustments). Recent advances in RL techniques (e.g., intrinsic reward-based exploration via random network distillation~\cite{burda19}) can further improve an agent's ability to discover effective filtering policies in complex or hard-exploration scenarios. In this work, PPO is combined with a composite reward design to learn adaptive filtering policies that remain robust under changing and previously unseen noise conditions.

\section{Background and Related Work}
Classical adaptive filters such as LMS and RLS are gradient-based/recursive algorithms that update filter weights to minimize instantaneous error. Their performance degrades under nonstationary or impulsive noise due to reliance on fixed learning rates and assumptions of Gaussian noise statistics. Kalman filters offer recursive minimum-variance estimation in state-space models but rely on accurate process and measurement noise covariances; adaptive Kalman variants exist, but they often resort to heuristic tuning. Wiener filters, derived by solving the Wiener--Hopf equations for stationary signal and noise with known second-order statistics, provide the optimal linear filter in the mean-square sense~\cite{haykin13}. However, Wiener filters are not suitable for time-varying noise and require prior knowledge of signal and noise correlation functions, which is impractical in many real-time systems.

Recently, RL-based approaches have emerged to tackle adaptive filtering and noise reduction without explicit modeling. Oh \emph{et al.}~\cite{oh21} first showed that a model-free RL (Q-learning) agent can outperform classical channel estimators under uncertain conditions. Xie \emph{et al.}~\cite{xie23} developed an RL-driven fractional order filter that is robust against impulsive (non-Gaussian) noise. RL has also been combined with model-based filtering: Lin \emph{et al.}~\cite{lin24} and He \emph{et al.}~\cite{he23} introduced reinforcement learning to adapt Kalman filter parameters online in nonstationary environments, and Marino and Guglieri~\cite{marino24} integrated an RL strategy with a Kalman filter for improved drone navigation under noise. More recently, Luo \emph{et al.}~\cite{luo24} applied deep RL to generative fixed-filter active noise control, demonstrating the potential of RL in active noise cancellation. These studies collectively indicate that RL-based adaptive filters can yield significant performance gains in dynamic noise scenarios by continually learning and adjusting filtering policies. Moreover, improvements in RL algorithms themselves (such as enhanced exploration strategies~\cite{burda19}) are directly relevant to adaptive filtering, as better exploration enables an agent to find more effective filtering strategies even in complicated or highly nonstationary environments.

\section{Problem Formulation}
Let $x(t)$ denote the clean signal and $n(t)$ a stochastic, possibly nonstationary noise process. The observed signal is
\begin{equation}
    y(t) = x(t) + n(t),
\end{equation}
and the goal is to produce an estimate $\hat{x}(t)$ that minimizes distortion over time. We frame adaptive filtering as a Markov decision process (MDP):
\begin{itemize}
    \item \textbf{State}: a feature representation of the recent noisy signal together with the current baseline-filter output.
    \item \textbf{Action}: a continuous update applied to the adaptive filter parameters.
    \item \textbf{Reward}: a composite performance signal balancing reconstruction quality and temporal stability.
\end{itemize}
The agent learns a policy $\pi$ that maximizes long-term filtering performance under changing noise statistics.

\section{PPO Framework and Composite Reward}
\label{sec:reward}
PPO is well suited to adaptive filtering because its clipped surrogate objective~\cite{schulman17}
\begin{equation}
L_{\mathrm{CLIP}}(\theta) = \mathbb{E}_t\!\Big[ \min\big( r_t(\theta) A_t,\; \mathrm{clip}(r_t(\theta), 1-\epsilon, 1+\epsilon) A_t \big) \Big]
\end{equation}
constrains policy changes and improves robustness when the reward landscape changes with the noise process. PPO also supports continuous action spaces, making it natural for coefficient-update policies.

We use the composite reward that combines three complementary objectives
\begin{equation}
R(t) = \alpha\, \Delta \mathrm{SNR}(t) - \beta\, \mathrm{MSE}(t) - \gamma\, \mathcal{S}(t),
\end{equation}
where $\Delta \mathrm{SNR}(t)$ is the improvement in signal-to-noise ratio, $\mathrm{MSE}(t)$ is the reconstruction error, and $\mathcal{S}(t)$ is a residual smoothness penalty. Where it penalizes temporal changes in the residual correction between consecutive steps, encouraging stable coefficient updates and suppressing oscillatory behavior, similar to a total variation, but not TV in signal processing sense.

During training, the MSE term is computed using clean reference signals available in simulation or labeled evaluation data. However, the PPO agent does \emph{not} receive the clean signal as part of its observation. Its state is constructed from the noisy input window together with the output of a baseline adaptive filter. Thus, the current framework is best interpreted as an offline-trained, online-deployed adaptive filtering approach: reference signals are used to shape the reward during training, but the learned policy runs without them at inference time.

\paragraph{Policy architecture}
The PPO agent uses an actor-critic architecture with a recurrent policy network. The actor is implemented as a single-layer LSTM with hidden size 128 that processes the observation sequence and outputs the mean of a Gaussian action distribution through a linear projection layer. The policy variance is parameterized by a learnable log-standard deviation vector shared across actions. Actions correspond to continuous coefficient updates applied to the adaptive FIR filter. The critic is implemented as a feed-forward network consisting of a linear layer with 128 hidden units, a ReLU activation, and a final linear layer producing the state-value estimate.

\paragraph{Training details and stability}
Training uses the Adam optimizer with learning rate $3\times10^{-4}$, discount factor $\gamma=0.99$, clipping parameter $\epsilon=0.2$, and 10 optimization epochs per policy update. Gradient norms are clipped to 0.5, and numerical safeguards are applied through observation and variance clamping. The recurrent architecture allows the policy to exploit temporal correlations in the signal, which is important for nonstationary noise processes.

\section{Experimental Setup}
\subsection{Synthetic signals}
We first evaluate the proposed method on a synthetic denoising task. The clean signal $x(t)$ is a sum of time-varying sinusoids with drifting amplitudes and multiple harmonics over 2048 samples.
\[
x(t) = \sum_{i=1}^{N} A_i(t) \cdot \sin(2\pi f_i t + \phi_i)
\]
Additive noise is drawn from Gaussian, Laplacian, impulse, pink, brown, and uniform distributions. PPO is trained only on Gaussian noise and evaluated on all noise types to test generalization. The agent observes a sliding window of recent noisy samples together with the corresponding baseline-filter output. Actions are coefficient updates applied to a finite impulse response (FIR) filter. We compare PPO with LMS, RLS, Wiener, and Kalman filters, each tuned by grid search.

The advantage is computed using baseline-subtracted returns:
\[
A_t = R_t - V(s_t), \quad \text{with} \quad R_t = \sum_{l=0}^{T-t-1} \gamma^l r_{t+l}
\]
where $V(s_t)$ is the value network’s estimate. Generalized advantage estimation (GAE) is omitted for simplicity and stability.

\subsection{Real ECG dataset}
To evaluate performance on real-world signals, we use the MIT-BIH Noise Stress Test Database (NSTDB) available through PhysioNet~\cite{goldberger2000physiobank,moody1984noise}. The dataset provides clean ECG recordings together with standardized physiological noise sources including baseline wander (bw), electrode motion (em), and muscle artifact (ma). These disturbances represent common artifacts encountered in mobile ECG monitoring. Following standard ECG denoising practice, noisy signals are generated by combining clean ECG recordings with the corresponding NSTDB noise sources. This yields a controlled but realistic evaluation setting with real physiological morphology and realistic artifact structure. The PPO policy is trained on ECG recordings with corresponding noise and then evaluated on other recordings of ECG disturbances without additional fine-tuning.

\section{Results}

\subsection{Synthetic generalization}
\begin{figure}[t]
\centering
\includegraphics[width=0.7\linewidth]{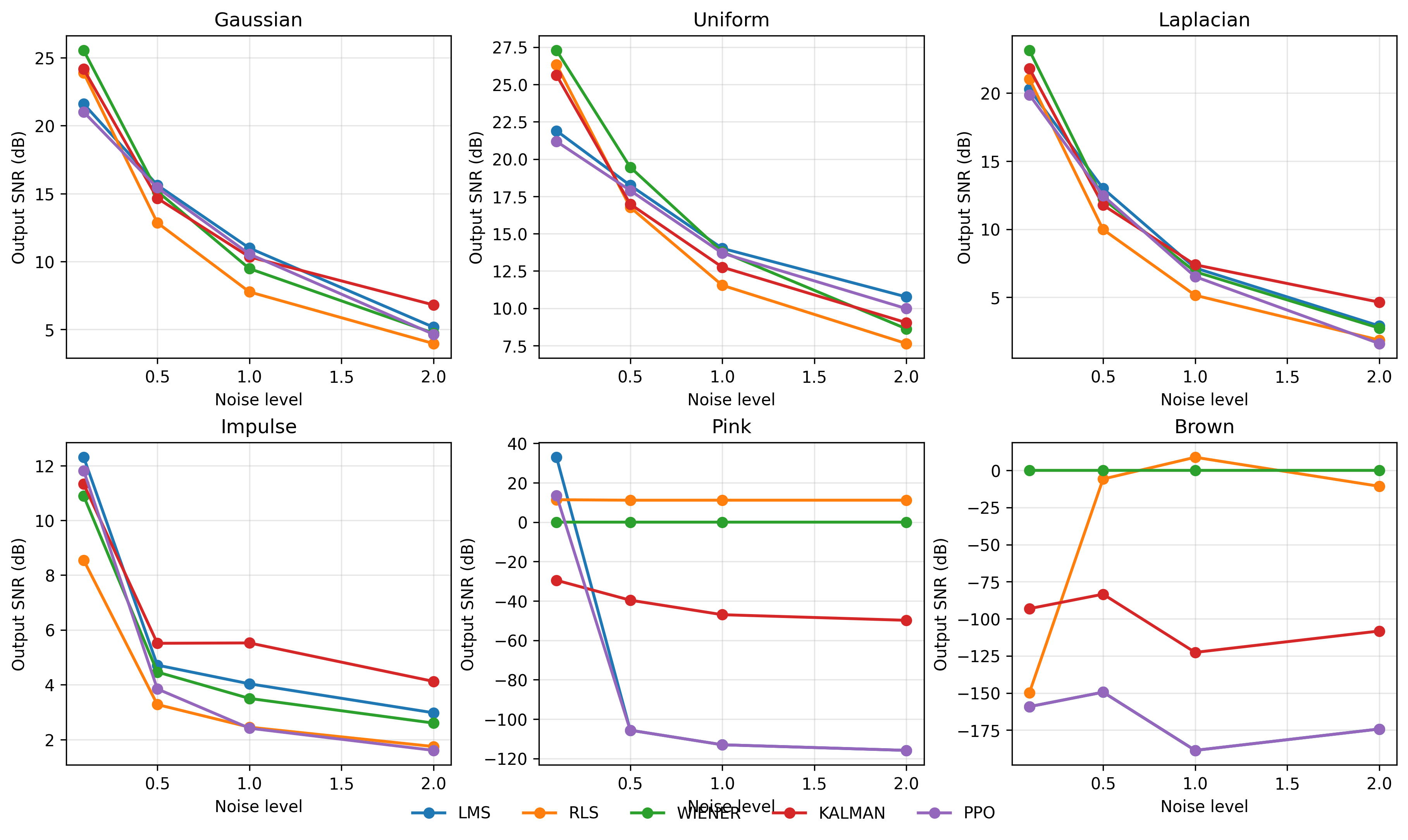}
\caption{Synthetic noise generalization across six noise families.}
\label{fig:synthetic_generalization}
\vspace{-5mm}
\end{figure}

Figure~\ref{fig:synthetic_generalization} evaluates generalization
to unseen noise families. PPO is trained only on Gaussian noise but
remains competitive under uniform, Laplacian, and impulse noise.
Pink and Brown noise are more challenging for all methods due to
their strong low-frequency structure.

\subsection{Real ECG performance}
Table~\ref{tab:snr_ecg} reports mean held-out SNR on real ECG data. PPO achieves the highest SNR across all three ECG noise conditions, slightly outperforming RLS while clearly exceeding LMS, Wiener, and Kalman.

\begin{table}[t]
\centering
\begin{minipage}[t]{0.47\linewidth}
\centering
\caption{Mean held-out SNR (dB) on real ECG signals.}
\label{tab:snr_ecg}
\small
\begin{tabular}{lccc}
\toprule
Method & BW & EM & MA \\
\midrule
LMS & 12.2 & 10.6 & 9.4 \\
RLS & 14.1 & 12.6 & 10.4 \\
Wiener & 7.3 & 6.9 & 7.2 \\
Kalman & 5.6 & 5.5 & 5.9 \\
\textbf{PPO} & \textbf{14.4} & \textbf{13.2} & \textbf{10.8} \\
\bottomrule
\end{tabular}
\end{minipage}\hfill
\begin{minipage}[t]{0.50\linewidth}
\centering
\caption{Impact of reward components on PPO filtering performance.}
\label{tab:ablation}
\small
\begin{tabular}{lccc}
\toprule
Reward & SNR & Gen. & Stab. \\
\midrule
Full & \textbf{16.6} & \cmark & \cmark \\
No smoothness & $<0$ & \xmark & \xmark \\
No MSE & 12.8 & \cmark & borderline \\
\bottomrule
\end{tabular}
\end{minipage}
\end{table}

\begin{figure}[t]
\centering
\includegraphics[width=\linewidth]{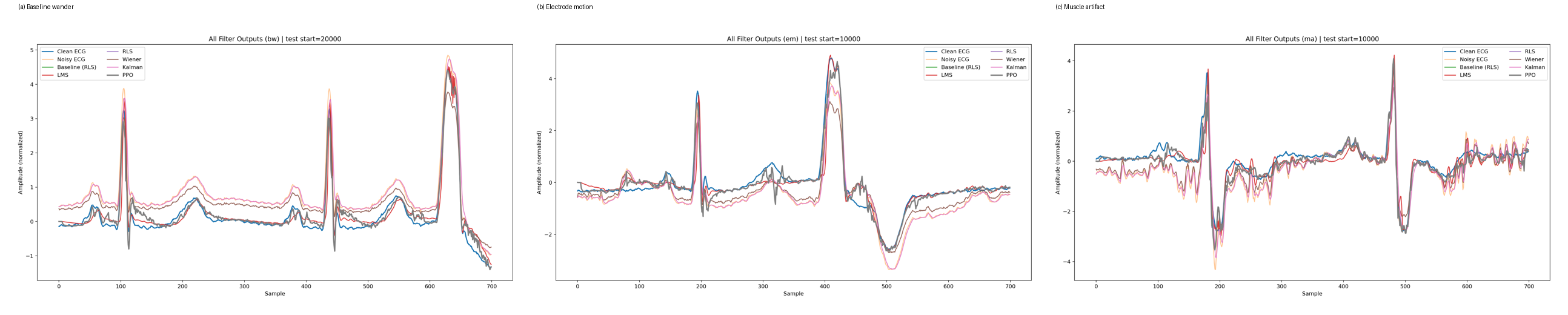}
\caption{Filtering performance on real ECG signals under baseline wander, electrode motion, and muscle artifact noise. PPO closely tracks the clean waveform and remains competitive with the strongest classical baseline, RLS, across all three settings.}
\label{fig:ecg_triptych}
\end{figure}

Figure~\ref{fig:ecg_triptych} provides qualitative ECG reconstructions. PPO consistently suppresses noise while preserving QRS morphology and larger-scale waveform shape. The gains are modest relative to RLS, but they are consistent across all three artifact classes, which is a more credible and practically meaningful result than a large single-noise improvement.

\subsection{Runtime on ECG data}
\begin{figure}[t]
\centering
\includegraphics[width=\linewidth]{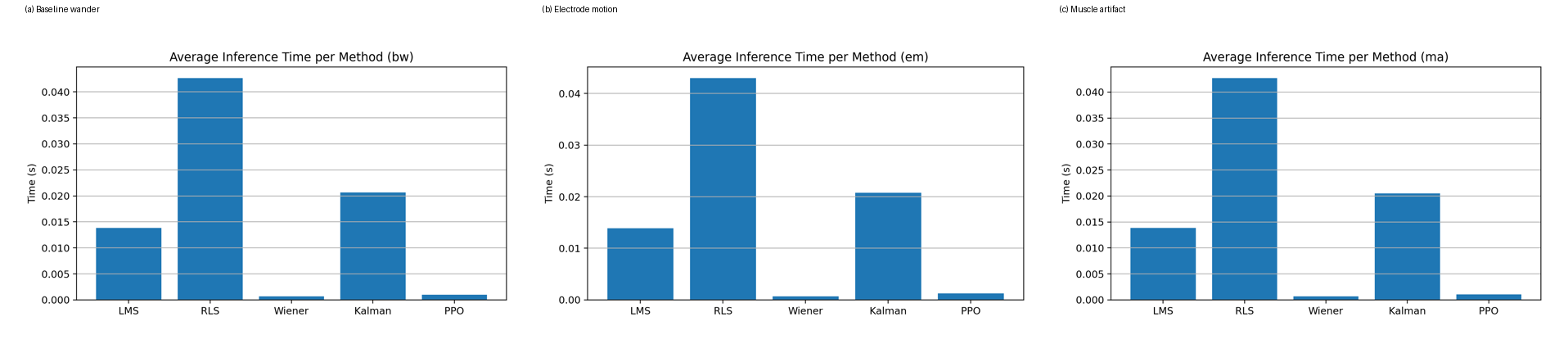}
\caption{Average inference time per method on ECG denoising tasks with baseline wander, electrode motion, and muscle artifact noise. PPO retains real-time performance while remaining markedly faster than RLS.}
\label{fig:runtime_ecg}
\end{figure}

In addition to denoising quality, we evaluate computational efficiency on the ECG setting. Figure~\ref{fig:runtime_ecg} shows average inference time per method for bw, em, and ma noise. PPO runs in approximately 1\,ms per inference, substantially faster than RLS and Kalman in our implementation and close to the cost of the lighter baselines. This supports the claim that PPO-based filtering remains feasible for real-time deployment after offline training.

\subsection{Training behaviour and stability}
\begin{figure}[t]
\centering
\begin{subfigure}{0.32\linewidth}
\includegraphics[width=\linewidth]{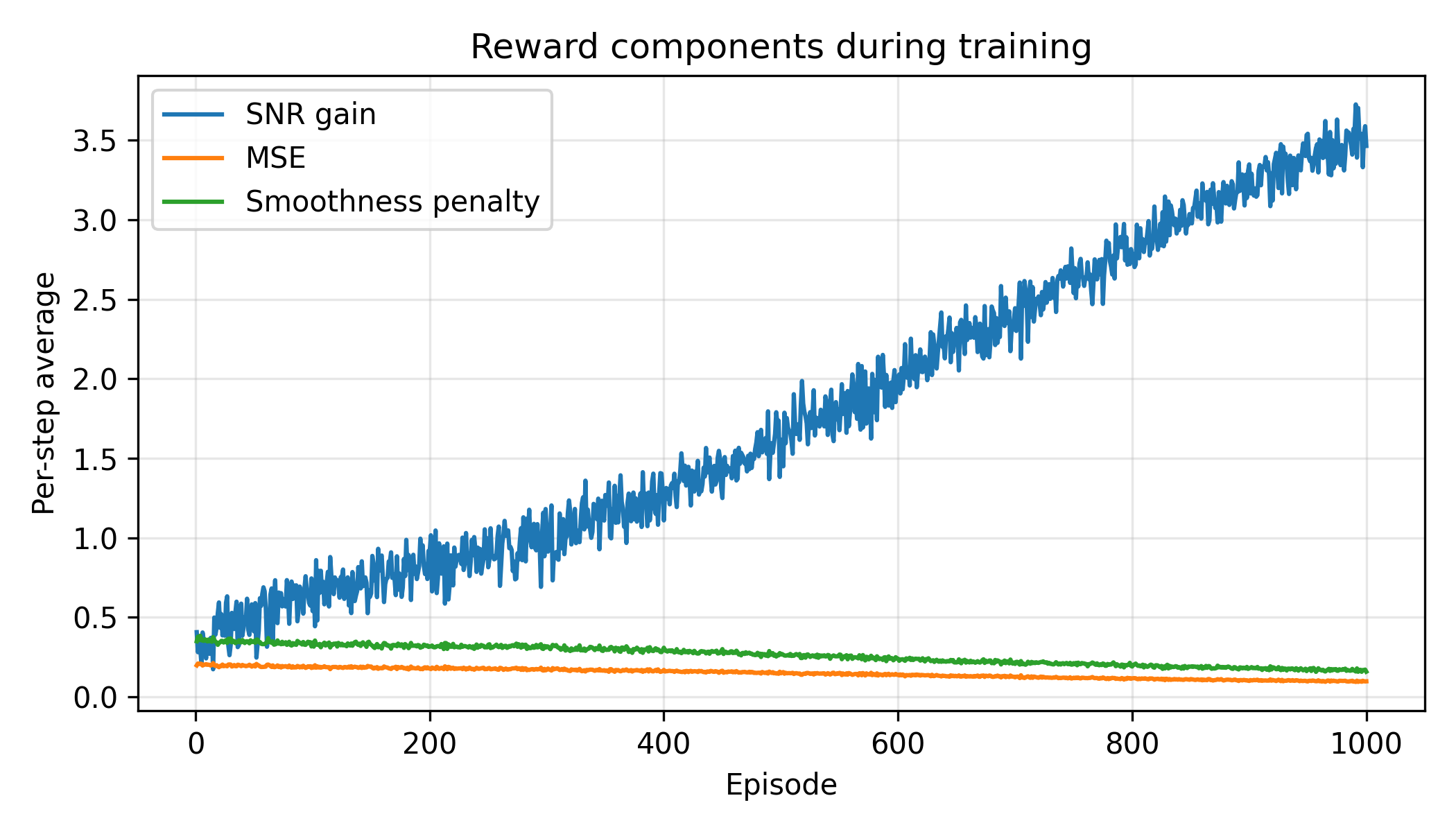}
\caption{Reward components}
\end{subfigure}\hfill
\begin{subfigure}{0.32\linewidth}
\includegraphics[width=\linewidth]{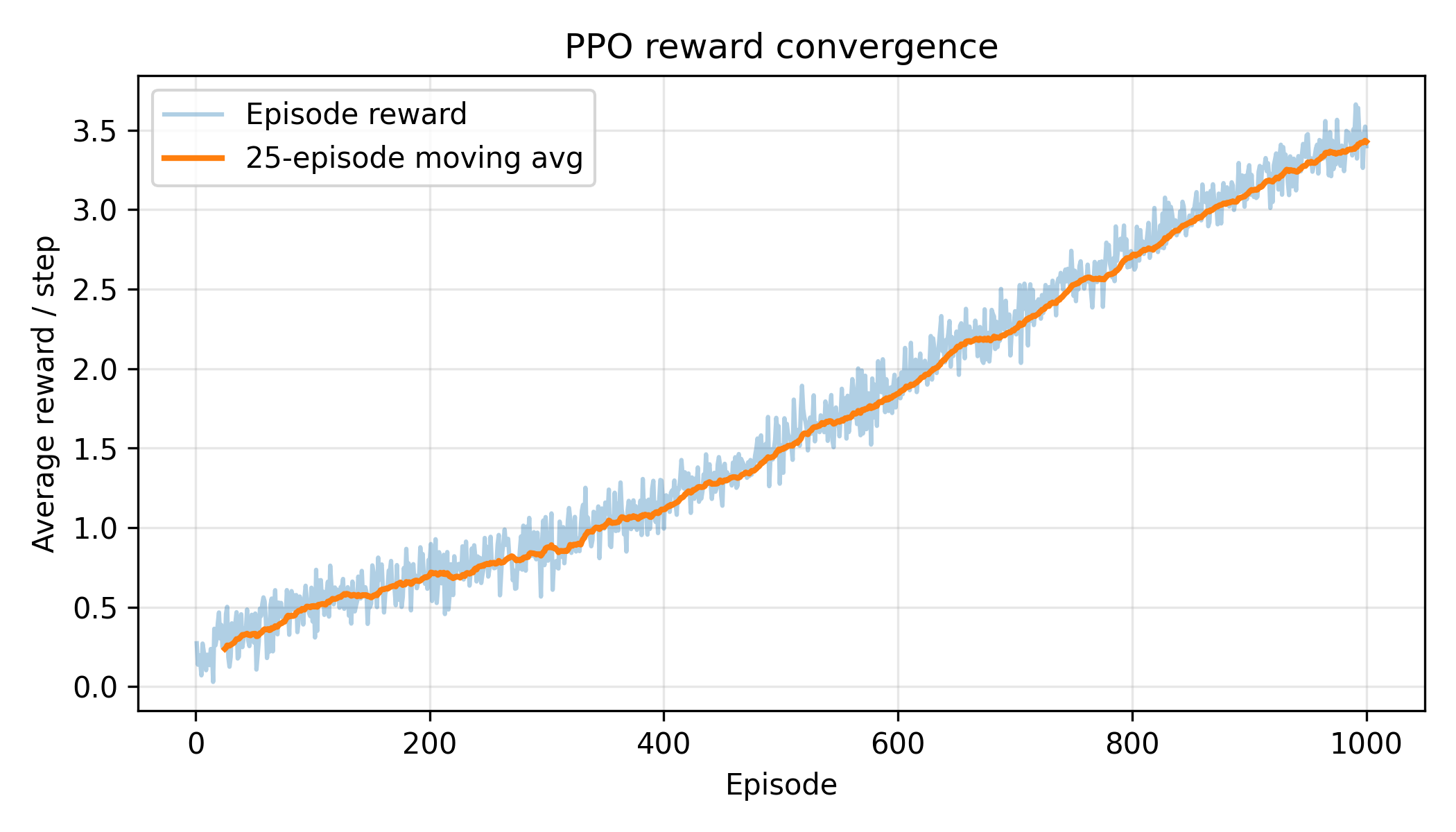}
\caption{Reward convergence}
\end{subfigure}\hfill
\begin{subfigure}{0.32\linewidth}
\includegraphics[width=\linewidth]{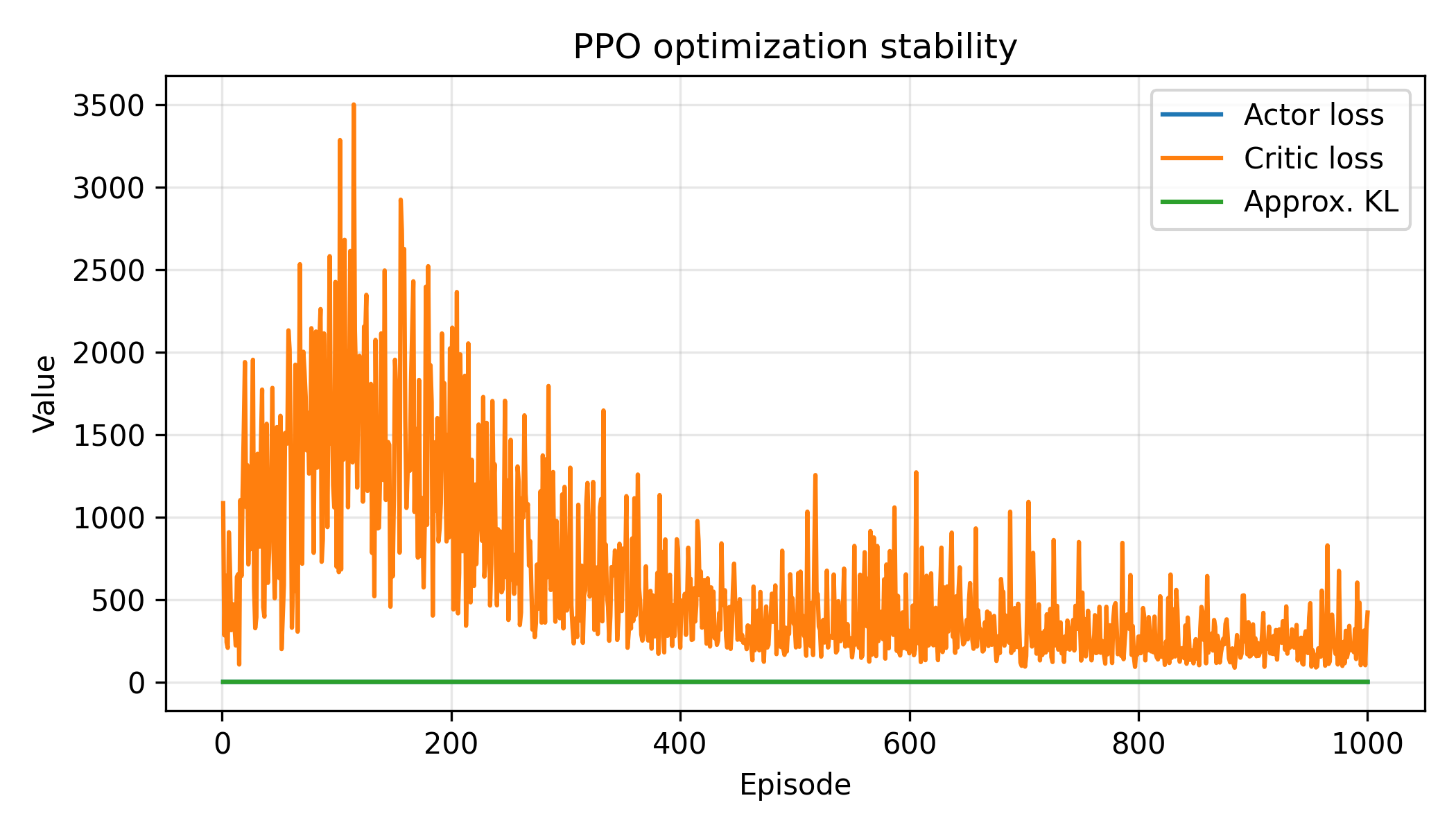}
\caption{Optimization stability}
\end{subfigure}
\caption{Training diagnostics for the PPO agent. The SNR gain increases over training, the MSE and smoothness penalties decrease, the episode reward improves steadily, and the actor/critic optimization remains stable after the initial exploration phase.}
\label{fig:training_diagnostics}
\end{figure}

Training was stable over 1000 episodes, and the composite reward was essential to that stability. Figure~\ref{fig:training_diagnostics} shows that the SNR gain term increases steadily during training while both the MSE term and the smoothness penalty decrease. The episode reward also improves consistently, and the actor/critic optimization dynamics stabilize after the initial exploration phase.

\subsection{Ablation study}
We performed controlled ablation experiments to isolate the effect of reward components. Results are summarized in Table~\ref{tab:ablation}. Here, \emph{Smoothness} denotes the temporal residual-change penalty.

These ablations confirm that removing either the MSE term or the smoothness component degrades stability and harms generalization, validating the need for the composite reward design.

\section{Discussion}
Experiments on real ECG signals provide insight on that, policies trained on synthetic mixtures or some split of real data still generalize and outperform on real biological signals, also clean reference signals are not directly shown to agent, and do not require them in inference time. Moreover, model's robustness on unseen noise indicates that the agent learns a transferable filtering policy rather than overfitting to a specific distribution. However, model requires a baseline model for path-finding, which might be mitigated with gamma parameter tuning. In that sense, the present framework is supervised in training but reference-free in deployment.

More broadly, the results show that PPO can learn coefficient-update policies that remain robust across multiple artifact types while preserving real-time inference speed. The remaining limitation is that training still depends on reference filters. Future work should therefore study unsupervised or self-supervised reward formulations and evaluate the same framework on additional biomedical and RF datasets.

\section{Conclusion}
This paper introduced a PPO-driven adaptive filtering framework based on a composite reward combining SNR improvement, MSE reduction, and residual smoothness. By formulating adaptive filtering as a Markov decision process, the agent learns coefficient-update policies that adapt to changing noise conditions rather than relying on fixed analytical update rules.

Experiments on synthetic signals and real ECG recordings show that PPO-based filtering generalizes beyond its training distribution, achieves the highest SNR among the tested methods on ECG denoising, and retains real-time inference cost. These findings support reinforcement learning as a practical tool for adaptive signal processing in dynamic and nonstationary environments.

\end{document}